\documentstyle[floats,aps,preprint]{revtex}
\begin{document}
\draft 
\preprint{\vbox{{\hbox{SOGANG-HEP 281/01}
                 \hbox{hep-th/0105112}}}}
\title{Quasinormal Modes and Choptuik Scaling in the Near Extremal
Reissner-Nordstr{\"o}m Black Hole}
\author
{Won Tae Kim\footnote{electronic address:wtkim@ccs.sogang.ac.kr}
and John J. Oh\footnote{electronic address:john5@string.sogang.ac.kr}}
\address{Department of Physics and Basic Science Research Institute,\\
         Sogang University, C.P.O. Box 1142, Seoul 100-611, Korea}
\date{\today}
\maketitle
\bigskip
\begin{abstract}
We study quasinormal modes of massless scalar and fermion fields
in the near extremal Reissner-Nordstr{\"o}m 
black hole, and relate them to Choptuik scaling form
following a recently proposed analytic approach. 
For both massless cases, quasinormal modes are shown to be proportional
to the black hole horizon and the Hawking temperature,
and the critical exponents are the same, although for the fermionic case there are two possible discrete quasinormal modes.
In addition, the critical exponent of 
the massive boson is also equivalent to that of the massless case.   
Finally, we discuss quasinormal modes and critical exponents in the
other models, and obtain some different critical exponents between
massless boson and massive one.
\end{abstract}
\pacs{PACS : 04.62.+v, 04.70.-s}
\bigskip

\newpage
One of the most exciting studies in gravitation
is the gravitational collapse of infalling matters.
When smooth initial data are slightly greater than
those of critical ones, then the black hole formation
appears, otherwise the matter fields eventually scattered below the
critical values of initial data \cite{cho}.
The dynamical behaviors are quite same and indistinguishable
after certain time, and then the final state of the collapsing matter
is determined by the initial data \cite{gun}. 

Since Ba${\tilde{\rm n}}$ados, Teitelboim, and Zanelli(BTZ) 
have found a nontrivial black hole solution in (2+1)-dimensional
anti-de Sitter(AdS) spacetime which has
a negative constant scalar curvature
\cite{btz}, it has been widely studied
in diverse branches of theoretical physics, including gravity and
string theory.
Recently, in connection with a critical behavior of black hole
formation in (2+1)-dimensional AdS spacetimes, 
several works have been done not only in
numerical ways but also analytical ones.
For a massless scalar field, critical behavior has
been studied in Ref. \cite{pc} by supposing a scaling relation of an AdS
black hole, 
\begin{equation}
  \label{scaling}
  M = K(p-p^{*})^{2\gamma},
\end{equation}
where the critical exponent yields $\gamma = 1.2 \pm 0.05$
by some numerical calculations.
An analytical approach \cite{gar} 
gives an exact critical solution for
a scalar field, which is compatible with the result obtained in Ref. \cite{pc}.
By a different numerical way, the critical collapse of scalar field in
(2+1)-AdS spacetimes has been studied and obtained the critical
exponent for a black hole formation, $\gamma = 0.81$ \cite{ho}.
In the case of dust-shell collapse, 
the exponent $\gamma=1/2$ has been obtained \cite{ps}.
Especially the BTZ black
hole formation has been studied in terms of group generators 
of AdS$_3$ isometries
using Gott time machine condition, and obtained a critical
exponent $\gamma=1/2$ for two colliding particles \cite{bs}.
 
On the other hand, a few works on quasinormal modes of black holes
show that quasinormal frequencies may be related to the critical
exponents in some special spacetime geometries. The quasinormal modes
of minimally coupled scalar fields in AdS-Schwarzschild black hole
have been studied in three \cite{cl}, four \cite{hh,wangs}, five, and seven dimensions \cite{hh}. And a
conjecture between quasinormal frequencies and a critical exponent,
$\omega \sim r_{H}/\gamma$, has been given \cite{hh}, and analytically studied 
in BTZ black hole background for a massive scalar field in Ref. \cite{bir}.

In this paper, we shall study some field equations on 
the near extremal RN black hole background which is effectively
described by AdS$_2$ black holes by assuming a spherical symmetric
dimensional reduction. Quasinormal modes of massless scalar and
fermionic fields in the near extremal RN black hole will be calculated following
Ref. \cite{bir}. For both cases, quasinormal modes are shown to be
proportional to the black hole horizon and the Hawking temperature,
and the critical exponents $\gamma_{b}$, $\gamma_{f}$ for boson and
fermionic respectively are all the same,
$\gamma_{b}=\gamma_{f}=1/2$. Furthermore, the critical exponent of
the massive boson is calculated, which turns out to be equivalent to
that of the massless case. Finally,
we shall discuss quasinormal modes and critical exponents in the
Jackiw-Teitelboim(JT) model \cite{jt}, which gives some different critical
exponents between massless boson and massive one.

We start with the Einstein-Maxwell system with a classical matter
in four dimensions, which is given by
\begin{equation}
  \label{ein-max}
   S_{\rm total} = S_{\rm EM} + S_{\rm M} = \frac{1}{16\pi G} \int d^4x\sqrt{-g_{(4)}}\left[R_{(4)} -
  \frac{1}{4} F^{\mu\nu}F_{\mu\nu} \right] + S_{\rm M},
\end{equation}
where $F$ is a field strength satisfying $F=dA$ and $S_{M}$ is a
matter action. The Einstein-Maxwell
action $S_{\rm EM}$ can
be reduced to the two-dimensional action,
\begin{equation}
  \label{2daction}
   S = \frac{1}{2\pi}\int dx^2 \sqrt{-g}e^{-2\phi}\left[R +
  2(\nabla \phi)^2 + 2\lambda^2 e^{2\phi} - \frac{1}{4}F^2\right],
\end{equation}
by assuming the spherically symmetric metric ansatz, $(ds)_{(4)}^2 =
  (ds)^2 + (e^{-2\phi}/ \lambda^2) d\Omega_{(2)}^2$ and $4\lambda^2 G
  = 2\pi$. The equation of motion for the action is exactly soluble
  and its solution is described by the two-dimensional charged
  Schwarzschild black hole,
  \begin{equation}
    \label{cschwarz}
    (ds)^2 = -\left(1-\frac{2GM}{x} + \frac{{Q}^2
  G^2}{x^2}\right) d^2t+\left(1-\frac{2GM}{x} + \frac{Q^2
  G^2}{x^2}\right)^{-1} d^2x,
  \end{equation}
with the dilaton and gauge field solutions, 
\begin{eqnarray}
  \label{dilgauge}
  & & \frac{e^{-2\phi}}{\lambda^2} = x^2, \nonumber \\
  & & e^{-2\phi} F^{\mu\nu} = \pi {Q} \epsilon^{\mu\nu},
\end{eqnarray}
where ${Q}$ is a electric charge of the black hole. In terms of the
inner and outer horizons $x_{\pm} = GM \pm G\sqrt{M^2 - {Q}^2}$,
Eq. (\ref{cschwarz}) can be rewritten as
\begin{equation}
  \label{csch}
  (ds)^2 = - \frac{(x-x_{+})(x-x_{-})}{x^2} d^2t + \frac{x^2 d^2x}{(x-x_{+})(x-x_{-})}.
\end{equation}
Then the Hawking temperature is 
\begin{equation}
  \label{hawk}
  T_{H} = \frac{\kappa_{+}}{4\pi}= \frac{(x_{+}-x_{-})}{4\pi x_{+}^2},
\end{equation}
where $\kappa_{+}$ is a surface gravity at the event horizon.

We are now interested in the near extremal case of the metric
(\ref{csch}). After performing the coordinate change such as $x=
G\tilde{x} - 2\pi G^2 Q^2 T_{H} + x_{+}$ and $t = 1/G \tilde{t}$, and
we take a near extremal
limit $x_{+} \rightarrow x_{-}$ under the fixed $T_{H}$ and $G
\rightarrow 0$ \cite{ss}, then the metric (\ref{csch}) can be written
as an AdS$_2$ black hole form,
\begin{equation}
  \label{ads2}
  (ds)^2 = -\left(-\mu^2 +
    \frac{\tilde{x}^2}{x_{+}^2}\right)d^2{\tilde{t}} + \left(-\mu^2 +
    \frac{{\tilde{x}^2}}{x_{+}^2}\right)^{-1} d^2{\tilde{x}},
\end{equation}
where $\mu^2 = (2\pi {Q} T_{H})^2$. 
By performing one more coordinate transformation 
such as $ \tilde{x} = (2x_{+} G
r - x_{+}^2)/(2x_{+}G)$, and taking limits as $G\rightarrow 0$ and $x_{+}\rightarrow x_{-}$ under the
fixed Hawking temperature, we obtain the JT model \cite{jt} upto the
leading order of a dilaton field, 
\begin{eqnarray}
  \label{jtsol}
  & & (ds)^2 = - \left( -\mu^2 + \frac{r^2}{x_{+}^2} \right) d^2t +
  \left(-\mu^2 + \frac{r^2}{x_{+}^2} \right)^{-1} d^2r + {\cal
  O}(r^3), \nonumber\\
  & & e^{-2\phi} = \pi x_{+} r.
\end{eqnarray}
So, the near extremal RN black hole can be effectively described
by the JT model \cite{fsn}. 
Note that we assume $G\rightarrow 0$, and this gives a large
mass black hole for a finite outer horizon $x_{+}$, since $GM \sim
x_{+} \sim GQ$ in the near extremal limit. Therefore, as the
black hole mass becomes large, the black hole
charge also becomes large. In Eq. (\ref{jtsol}), since $\mu \sim
1/G^2 x_{+}$ as $G\rightarrow 0$, the horizon of AdS black hole
becomes large in the near extremal limit. The scalar curvature for
Eq. (\ref{cschwarz}) is  
\begin{equation}
  \label{scrn}
  R = \frac{6{Q}^2 G^2 - 4GMx}{x^4},
\end{equation}
and in an appropriate limit of the near extremal black hole, the given
scalar curvature approaches to the constant, 
$R \propto - 1/{x_{+}^2}$.

Next, we shall study quasinormal modes of some
matter fields in this near extremal RN black hole in the spherical
symmetric reduction, and relate
quasinormal modes to the critical exponent of scaling law for the black
hole formation.
So the massless bosonic and fermionic fields in four dimensions are
first considered, which is effectively described by the following two-dimensional
action,
\begin{equation}
  \label{matt2}
  S_{M} = \frac{1}{2\pi} \int dx^2 \sqrt{-g} e^{-2\xi\phi}
  \left[-\frac{1}{2} (\nabla f)^2\right],
\end{equation}
by taking only s-wave sector of matter fields. Note that for $\xi=1$, it
describes a bosonic field in four dimensions, while the fermionic case
is for $\xi=0$ which is written after bosonization \cite{as0}. 
The equation of motion from the action (\ref{matt2}),
\begin{equation}
  \label{eqnmot}
  \Box f - 2 \xi \partial_{\mu}\phi \partial^{\mu} f = 0,
\end{equation}
can be solved under the dilatonic AdS$_2$ black hole background (\ref{jtsol})
This metric has a new horizon at $r= r_{H}
\equiv x_{+}\mu$, and a new Hawking temperature $T^{\rm AdS}_{H}$ is
related to the original temperature as $T^{\rm AdS}_{H} = (Q/x_{+}) T_{H}$. 

In fact, quasinormal modes in AdS geometry has been defined in Refs. \cite{hh,bir}. In the near horizon limit, only ingoing waves can be
allowed while some fields should vanish at the AdS boundary since the potential diverges at infinity. This is a
prominent feature of quasinormal modes in AdS geometry which differs from
those in asymptotically flat one. So there exists a discrete
set of quasinormal modes $\omega$ with a negative imaginary part
\cite{hh}.
In the two-dimensional system with an AdS geometry, we expect the
solution of wave equation can be exactly solved and will be a form of
a hypergeometric function \cite{ko}. 

First, we consider the bosonic case of $\xi=1$ to solve the wave equation
(\ref{eqnmot}). Using the background metric (\ref{jtsol}) and the
separation of variables $f(r,t) = R(r) e^{-i\omega t}$, the equation of
motion can be written as
\begin{equation}
  \label{a1eqn}
  \left(r^2 - r_{H}^2\right) \partial_{r}^2 R(r) +
  \frac{1}{r} \left(3r^2 -r_{H}^2\right) \partial_{r} R(r) +
  \frac{\omega^2 x_{+}^4}{(r^2 -r_{H}^2)} R(r) =0,
\end{equation}
where $r_{H} = \mu x_{+}$.
By means of the coordinate change $z=(r^2 - r_{H}^2)/r^2$,
Eq. (\ref{a1eqn}) becomes  
\begin{equation}
  \label{hyper}
  z(1-z)\partial_{z}^2 R(z) + (1-z) \partial_{z} R(z) + \frac{A}{z} R(z) =0,
\end{equation}
where $A=\omega^2 x_{+}^4 /4r_{H}^2$. Note that a new variable $z$
spans from 0 to 1 as $r$ goes from $r_{H}$ to infinity. To remove a
singular point at $z=0$, we set $R(r) = z^{\alpha} g(z)$.
Then the equation of motion can be solved exactly, which yields a
solution of hypergeometric functions $F(a,b,c;z)$ and $z^{1-c}
F(a-c+1,b-c+1,2-c;z)$, and $\alpha^2$ is determined as $-A$, where
$a=b=\alpha$, and $c=1+2\alpha$.
So, we have a general solution from Eq. (\ref{a1eqn}),
\begin{equation}
  \label{a1sol}
  R(z) = C_{\rm out} z^{\alpha} F(\alpha, \alpha, 1+2\alpha;z) +
  C_{\rm in} z^{-\alpha} F(-\alpha, -\alpha, 1-2\alpha;z),
\end{equation}
where $C_{\rm in}$ and $C_{\rm out}$ are ingoing and outgoing
coefficients, respectively, and $\alpha = i\omega x_{+}^2 /
2r_{H}$. We take a plus sign of $\alpha$ due to $\alpha \rightarrow -
\alpha$ symmetry for convenience. 

Applying the quasinormal mode condition, $C_{\rm out} = 0$ at the horizon
as $z\rightarrow 0$, and we have only ingoing near horizon solution,
\begin{equation}
  \label{a1nearho}
  R(z) = C_{\rm in} z^{-\alpha} F(-\alpha, -\alpha, 1-2\alpha;z).
\end{equation}
To give a boundary condition at the AdS boundary, we take a transformation of
$z \rightarrow 1-z$. In this case, since we have $c=a+b+m$ with $m=1$,
we have to use the following transformation rule\cite{as},
\begin{eqnarray}
  \label{1ztrans}
  & &F(a,b,a+b+m;z) = \frac{\Gamma(m) \Gamma(a+b+m)}{\Gamma(a+m)
  \Gamma(b+m)} \sum_{n=0}^{m-1} \frac{(a)_{n} (b)_{n}}{n! (1-m)_{n}}
  (1-z)^{n}\nonumber \\
 & & \qquad\qquad - \frac{\Gamma(a+b+m)}{\Gamma(a)\Gamma(b)} (z-1)^{m}
  \sum_{n=0}^{\infty} \frac{(a+m)_{n}
  (b+m)_{n}}{n!(n+m)!}(1-z)^{n}\nonumber \\
 & & \qquad\qquad \times \left[ \ln (1-z) - \psi(n+1) -\psi(n+m+1) + \psi(a+n+m) +\psi(b+n+m)\right].
\end{eqnarray}
To vanish the solution at infinity($z\rightarrow 1$), we take a
condition of either $a+1 = -n$ or $b+1=-n$. It gives a discrete
quasinormal frequency $\omega$ as 
\begin{equation}
  \label{wavefre}
  \omega = - i\frac{2r_{H}}{x_{+}^2} (n+1),
\end{equation}
so that the critical exponent for the massless boson case is regarded as
$\gamma_{b}^{(0)} = 1/2$. In terms
of the Hawking temperature, Eq. (\ref{wavefre}) is written as
\begin{equation}
  \label{hawkfre}
  \omega = -i 4\pi \frac{{Q}}{x_{+}} T_{H} (n+1) = -i 4\pi T^{\rm AdS}_{H} (n+1).
\end{equation}
So this quasinormal mode is proportional to the horizon $r_{H}$,
and the Hawking temperature $T^{\rm AdS}_{H}$ of the AdS black hole
\cite{hh}.

In Eq. (\ref{wavefre}), the discretized quasinormal
frequency has a negative imaginary part. The most general proof of
this argument is given in Ref. \cite{hh}. It can be realized
by assuming the field equation under the gravitational background
with the boundary condition at the asymptotic region.
By using the metric in terms of Eddington-Finkelstein coordinates, $v = t + r_{*}= t + r/g(r)$ as
\begin{equation}
  \label{metedd}
  (ds)^2 = - g(r) d^2 v + 2 dv dr,
\end{equation}
where $g(r) = (\mu^2 - r^2 / x_{+}^2 )$ and performing separation
 of variables, $f(v,r) = R(r) e^{-i\omega v}$ with conditions $g(r_{H})
 =0$ and $R(\infty)=0$, the useful relation is given as \cite{hh}
\begin{equation}
  \label{proof}
  \int_{r_{H}}^{\infty} dr \left[ g(r) |\frac{dR}{dr}|^2 \right] = -
  \frac{|\omega|^2 |R(r_{H})|^2}{{\rm Im} \omega}.
\end{equation}
Therefore, quasinormal frequency has a negative imaginary part since
$g(r)$ is positive definite outside the horizon.

Now we consider the fermionic case of $\xi=0$. In this case, we have
somewhat different type of scalar field equation from
Eq. (\ref{a1eqn}) since there does not exist a nontrivial dilaton
coupling in the field equation, which is given as
\begin{equation}
  \label{a0eqn}
  (r^2 - r_{H}^2) \partial_{r}^2 R(r) + 2r \partial_{r} R(r) +
  \frac{\omega^2 x_{+}^4}{(r^2 - r_{H}^2)} R(r) = 0.
\end{equation}
Using the coordinate transformation $z=(r^2 - r_{H}^2)/{r^2}$ and
setting $R(z) = z^{\alpha} g(z)$ to remove a singularity at $z=0$, the
wave equation becomes
\begin{equation}
  \label{a0waveeq}
  z(1-z) \partial_{z}^2 g(z) + \frac{1}{2} \left[ 2+4\alpha -
  (3+4\alpha) z\right] \partial_{z} g(z) - \frac{1}{2} \alpha (2\alpha
  + 1) g(z) =0,
\end{equation}
where $\alpha^2$ is determined as $-A=-\omega^2 x_{+}^4 / 4
r_{H}^2$. The solution of Eq. (\ref{a0waveeq}) is given as 
\begin{equation}
  \label{a0sol}
  R(z) = D_{\rm in} z^{-\alpha} F\left(\frac{1}{2} - \alpha, -\alpha, 1-2
  \alpha;z\right) + D_{\rm out} z^{\alpha} F\left(\alpha, \frac{1}{2} + \alpha, 1+2\alpha;z\right),
\end{equation}
where $D_{\rm in}$ and $D_{\rm out}$ are ingoing and outgoing
coefficients, respectively, and $\alpha = i\omega x_{+}^2 /
2r_{H}$. From the definition of quasinormal modes, we set $D_{\rm out}
=0 $. For the present case, one should use a different transformation
rule of $z\rightarrow 1-z$ \cite{as}, 
\begin{eqnarray}
  \label{a0z1ztrans}
F(a,b,c;z) &=& \frac{\Gamma(c)\Gamma(c-a-b)}{\Gamma(c-a)\Gamma(c-b)}
F(a,b,a+b-c+1;1-z) \nonumber \\
&+& (1-z)^{c-a-b}\frac{\Gamma(c)\Gamma(a+b-c)}{\Gamma(a)\Gamma(b)} F(c-a,c-b,c-a-b+1;1-z),
\end{eqnarray}
where $a=1/2-\alpha$, $b=-\alpha$, and $c=1-2\alpha$.
To vanish the solution as $z\rightarrow 1$ gives a condition either
$c-a=-n$ or $c-b=-n$. This yields two types of discrete quasinormal
frequencies as
\begin{equation}
  \label{a0quasi}
  \omega = -i \frac{2 r_{H}}{x_{+}^2}\left(n+\frac{1}{2}\right),
\end{equation}
or
\begin{equation}
  \label{a0quasi2}
  \omega = -i \frac{2 r_{H}}{x_{+}^2} (n+1).
\end{equation}
As for the critical exponent for the massless fermionic case, it
yields $\gamma_{f}^{(0)}=1/2$, which is the same value with that of the
bosonic case.

Now let us consider that a massive scalar field in the s-wave sector. It
can be reduced to the two-dimensional massive scalar field equation in the
dilatonic AdS black hole background,
\begin{equation}
  \label{massive}
  \Box f - 2\partial_{\mu} \phi \partial^{\mu} f - m^2 f = 0.
\end{equation}
Using the separation of variables $f(r,t) = R(r)e^{-i\omega t}$ and coordinate
change $z=(r^2-r_{H}^2)/r^2$ for $0 \le z \le 1$, the wave equation
becomes
\begin{equation}
  \label{mawe}
  z(1-z)\partial_{z}^2 R(z) + (1-z)\partial_{z} R(z) + \left[
  \frac{\omega^2 x_{+}^4}{4r_{H}} \left(\frac{1}{z}\right) - \frac{m^2
  x_{+}^2}{1-z} \right] R(z) =0.
\end{equation}
To remove two singular points at $z=0$ and $z=1$, one can redefine $R(z)
= z^{\alpha}(1-z)^{\beta} g(z)$, Eq. (\ref{mawe}) becomes a standard
form of a differential equation satisfying a hypergeometric function,
\begin{equation}
  \label{eqnhyper}
  z(1-z)\partial_{z}^2 g(z) + \left[1+2\alpha - (1+2\alpha + 2\beta)
  z\right] \partial_{z} g(z) - (\alpha + \beta)^2 g(z) = 0,
\end{equation}
where $\alpha = i \omega x_{+}^2 /2r_{H}$ and $\beta = (1-\sqrt{1+4m^2 x_{+}^2})/2$.
The general solution of Eq. (\ref{eqnhyper}) is
\begin{eqnarray}
  \label{gensol}
  R(z) &=& E_{\rm out} z^{\alpha} (1-z)^{\beta} F(\alpha+\beta, \alpha+\beta,
  1+2\alpha;z)\nonumber \\
       & &\qquad + E_{\rm in} z^{-\alpha} (1-z)^{\beta} F(-\alpha+\beta,
  -\alpha+\beta, 1-2\alpha;z),
\end{eqnarray}
and $E_{\rm out}=0$ by the definition of quasinormal
modes. At this stage, we use the transformation rule (\ref{a0z1ztrans}) to make
the solution vanish at infinity, and this gives a quasinormal frequency 
\begin{equation}
  \label{massquasi}
  \omega = -i \frac{2r_{H}}{x_{+}^2} \left(n+\frac{1}{2} + \frac{1}{2}
  \sqrt{1 + 4m^2 x_{+}^2}\right),
\end{equation}
and it gives the critical exponent for the massive boson,
$\gamma_{b}^{(m)} = 1/2$.

We have studied quasinormal modes and critical exponents of classical matters
including massless boson and fermion, and massive bosonic field of
s-wave sector in the near extremal RN black
hole. Quasinormal frequencies are obtained by solving the wave
equation from the definition of quasinormal modes. Following the
relation $\omega \sim r_{H} / \gamma$ in Ref. \cite{bir}, as a result,
the critical exponents could be obtained as $\gamma_{b}^{(0)} =
\gamma_{f}^{(0)} = \gamma_{b}^{(m)} = 1/2$ for massless scalar,
massless fermion, and massive scalar fields. 

Note that the charge scaling relation was not derived here because
the black hole charge has been regarded as a fixed value in order to
use AdS geometry and the useful relation to quasinormal modes. 
On the other hand, the charge scaling and its universality 
in critical collapse of charged scalar fields
have been studied in
Ref. \cite{gm}, and a mass scaling, $M \sim
(p-p_{*})^{\gamma}$, and a new scaling with
respect to the black hole charge, $Q\sim (p-p_{*})^{\delta}$, have
been obtained where
the corresponding exponents, $\gamma=0.374\pm 0.001$ and
$\delta=0.883\pm0.007$, respectively.
However, our approach is valid
only for large black holes, since the RN black hole can be described
by the dilatonic AdS$_2$ black hole which behaves as a large black hole
in the near extremal limit. 
For these large black holes, it can be shown that
quasinormal modes scale with the black hole temperature $T^{\rm
  AdS}_{H}$. These will decay exponentially, which translates into a
time scale for the approach to the thermal equilibrium. Time scale is
given by the imaginary part of the lowest quasinormal mode $\tau
=1/\omega_{I}$, where $\omega = \omega_{R} - i \omega_{I}$. These time
scales have universal values in that all scalar fields will decay at
these rates \cite{hh}. For our cases, the time scale of the lowest
quasinormal mode can be obtained for $n=0$ and inversely proportional
to the Hawking temperature, 
\begin{equation}
  \label{timescale}
  \tau \sim \frac{x_{+}^2}{r_{H}} = \frac{1}{T^{\rm AdS}_{H}}.
\end{equation}
This means that the more black hole radiates, the faster it decays and
approaches to the thermal equilibrium. For a massless bosonic
case($\xi=1$), we have a time scale $\tau_{b}^{(0)} = 1/4\pi T^{\rm
  AdS}_{H}$ while the time scale of a massive boson is given as
\begin{equation}
  \label{masstscale}
  \tau_{b}^{(m)} = \frac{1}{2\pi T_{H}^{\rm AdS}} \left(1+\sqrt{1+4m^2 x_{+}^2}\right)^{-1}.
\end{equation}
Note that the fluctuated black hole from perturbations of the massive
boson will go to the thermal equilibrium faster than those of the massless
bosonic case since $\tau_{b}^{(0)} > \tau_{b}^{(m)}$.
For the fermionic case, two kinds of time scale, $\tau_{f}^{(0)} =
1/4 \pi T^{\rm AdS}_{H}$ and $\tau_{f}^{(0)} = 1/2\pi T^{\rm
  AdS}_{H}$, are possible due to Eqs. (\ref{a0quasi}) and (\ref{a0quasi2}). 

On the other hand, one might be interested in the critical exponents of 
two-dimensional scalar fields instead of the above 
dimensionally
reduced four-dimensional ones on this dilatonic AdS$_{2}$ black hole
background of the JT model \cite{jt}, 
which is just the two-dimensional version of the 
initial work done in the BTZ black hole \cite{bir}. 
For the case of the 
two-dimensional massless boson, 
it is just what we have already done as far as
we remove the dilaton coupling in Eq. (\ref{eqnmot}), which
corresponds to the equation of motion for the spherical symmetric
fermion in the four dimensions. Therefore, the relation for the
massless field in two dimensions is $\omega = - i 2r_{H}/x_{+}^2 (n+1)$ and
$\gamma_{{\rm massless}}^{\rm AdS_2}=1/2$. Then, what about the
massive scalar field in the AdS$_{2}$ black hole background? The above
similar calculations can be straightforwardly done, by dropping the
dilaton coupling in Eq. (\ref{eqnmot}) and adding mass
term. Performing some change of variables as $z=(r-r_{H})/(r+r_{H})$
and imposing the above quasinormal condition, the following relation
is obtained as $\omega = -i r_{H}/x_{+}^2 (n+1)$. It gives
interestingly twice value compared to that of the massless case,
$\gamma_{{\rm massive}}^{\rm AdS_2}=1$. This is more or less
surprising in that the critical exponent of the massive scalar field
is different from
that of the massless one in the AdS$_{2}$ black hole, whereas this is not
the case for the BTZ black hole i.e., the critical exponents are the
same, $\gamma^{\rm BTZ}_{\rm massive}=\gamma^{\rm BTZ}_{\rm
  massless}=1/2$ \cite{bir}.  

The final comment is that we have not justified 
the scaling relation (\ref{scaling}). Of course, in the
three-dimensional BTZ case, it is proved in the numerical way
in Refs. \cite{pc}. So, it would be interesting to find out how
to obtain the squared scaling relation in two-dimensional AdS.

\vspace{1cm}

{\bf Acknowledgments}\\
This work was supported by grant No. 2000-2-11100-002-5
from the Basic Research Program of the Korea Science and
Engineering Foundation. We would like to thank to P. P. Jung for
helpful discussions.



\begin{references}
\bibitem{cho} M. W. Choptuik, {\it Universality and Scaling in
    Gravitational Collapse of a Massless Scalar Field},
    Phys. Rev. Lett. {\bf 70} (1993) 9.
\bibitem{gun} For a recent review, 
              C. Gundlach, {\it Critical phenomena in gravitational
              collapse}, {\tt [gr-qc/0001046]}.
\bibitem{btz} M. Ba${\tilde{\rm n}}$ados, C. Teitelboim, and
  J. Zanelli, {\it The Black Hole in Three-dimensional Space-time}, 
  Phys. Rev. Lett. {\bf 69} (1992) 1849, {\tt [hep-th/9204099]}.
\bibitem{pc} F. Pretorius and M. W. Choptuik, Phys. Rev. D62 (2000)
  124012, {\tt [gr-qc/0007008]}.
\bibitem{gar} D. Garfinkle, {\it An exact solution for 2+1 dimensional
  critical collapse}, Phys. Rev. {\bf D63} (2001) 044007, {\tt
  [gr-qc/0008023]}.
\bibitem{ho} V. Husain and M. Olivier, {\it Scalar field collapse in
  three-dimensional AdS spacetime}, Class. Quant. Grav. {\bf 18}
  (2001) L1, {\tt [gr-qc/0008060]}. 
\bibitem{ps} Y. Peleg and A. R. Steif, {\it Phase Transition for
  Gravitationally Collapsing Dust Shells in (2+1)-dimensions},
  Phys. Rev. {\bf D51} (1995) 3992, {\tt [gr-qc/9412023]}.
\bibitem{bs} D. Birmingham and S. Sen, {\it Gott Time Machines, BTZ
  Black Hole Formation, and Choptuik Scaling}, Phys. Rev. Lett. {\bf
  84} (2000) 1074, {\tt [hep-th/9908150]}.
\bibitem{cl} V. Cardoso and J. P. S. Lemos, {\it Scalar, Electromagnetic and Weyl Perturbations of BTZ black holes: Quasinormal Modes}, {\tt [gr-qc/0101052]}.
\bibitem{hh} G. Horowitz and V. E. Hubeny, {\it Quasinormal modes of
    AdS black holes and the approach to thermal equilibrium},
    Phys. Rev. {\bf D62} (2000) 024027, {\tt [hep-th/9909056]}.
\bibitem{wangs} B. Wang, C.-Y. Lin, and E. Abdalla, {\it Quasinormal modes of Reissner-Nordstrom Anti-de Sitter black holes}, Phys. Lett. {\bf B481} (2000) 79 ; J.-M. Zhu, B. Wang, and E. Abdalla, {\it Object picture of quasinormal ringing on the background of small Schwarzschild anti-de Sitter black holes}, Phys. Rev. {\bf D63} (2001) 124004.
\bibitem{bir} D. Birmingham, {\it Choptuik Scaling and Quasinormal
    Modes in the AdS/CFT Correspondence}, {\tt [hep-th/0101194]}.
\bibitem{jt} R. Jackiw, in Quantum Theory of Gravity, edited by
    S. M. Christensen (Hilger, Bristol, 1984); C. Teitelboim, in
    Quantum Theory of Gravity, edited by S. M. Christensen (Hilger,
    Bristol, 1984).
\bibitem{ss} M. Spradlin and A. Strominger, {\it Vacuum States for
    AdS$_2$ Black Holes}, J. High Energy Phys. {\bf 9911} (1999) 021, {\tt
    [hep-th/9904143]}.
\bibitem{fsn} A. Fabbri, D. J. Navarro, and J. Navarro-Salas, {\it Quantum
    evolution of near-extremal Reissner-Nordstr{\"o}m black holes},
  Nucl. Phys. {\bf B595} (2001) 381, {\tt[hep-th/0006035]}.
\bibitem{as0} M. G. Alford and A. Strominger, {\it S Wave Scattering of
    Charged Fermions by a Magnetic Black Hole}, Phys. Rev. Lett. {\bf
    69} (1992) 563, {\tt [hep-th/9202075]}.
\bibitem{ko} W. T. Kim and J. J. Oh, {\it Dilaton Driven Hawking
    Radiation in AdS$_2$ Black Hole}, Phys. Lett. {\bf B461} (1999)
    189, {\tt [hep-th/9905007]}.
\bibitem{as} M. Abramowitz and I. A. Stegun, {\it Handbook of
    Mathematical Functions}, Dover Publication Inc., New York, ninth
    printing, (1970).
\bibitem{gm} C. Gundlach and J. M. Martin-Garcia, {\it Charge
    scaling and universality in critical collapse}, Phys. Rev. {\bf
    D54} (1996) 7353, {\tt [gr-qc/9606072]}.
\end{references}
\end{document}